\documentclass[twocolumn,showpacs,preprintnumbers,amsmath,amssymb]{revtex4}
\usepackage{graphicx}
\usepackage{dcolumn}
\usepackage{bm}
\begin{document}

\title{Fast Quantum Algorithm for Numerical Gradient Estimation}
\author{Stephen P. Jordan}
\affiliation{MIT Physics, 77 Massachusetts Avenue, Cambridge,
  Massachusetts 02139}
\date{\today}

\newcommand{\ud}{\mathrm{d}}
\newcommand{\bra}[1]{\langle #1|}
\newcommand{\ket}[1]{|#1\rangle}
\newcommand{\braket}[2]{\langle #1|#2\rangle}
\newcommand{\bx}{\mathbf{x}}
\newcommand{\bd}{\boldsymbol{\delta}}
\newcommand{\bN}{\mathbf{N}}
\newcommand{\bk}{\mathbf{k}}
\newcommand{\bu}{\mathbf{u}}
\newcommand{\bo}{\mathbf{0}}
\newcommand{\Ket}[1]{\left|#1\right>}

\begin{abstract}
Given a blackbox for $f$, a smooth real scalar function of $d$
real variables, one wants to estimate $\nabla f$ at a given point
$\bx =\left(x_1,x_2,\ldots,x_d\right)$ with $n$ bits of
precision. On a classical computer this requires a minimum of
$d+1$ blackbox queries, whereas on a quantum computer it requires
only one query regardless of $d$. The number of bits of precision to
which $f$ must be evaluated matches the classical requirement in the
limit of large $n$.
\end{abstract}

\maketitle

In the context of many numerical calculations, blackbox query
complexity is a natural measure of algorithmic efficiency. For
example, in optimization problems, the function evaluations are
frequently the most time consuming part of the computation, and an
efficient optimization algorithm is therefore one which uses as few
function evaluations as possible \cite{nr}.

Here we investigate the query complexity of numerically estimating the
gradient of a blackbox function at a given point. We find that
gradients can be estimated on a quantum computer using a single
blackbox query. The algorithm which achieves this can be viewed as a
generalization of the Bernstein-Vazirani \cite{Bernstein} algorithm,
which has been described in other contexts \cite{Mosca, Hoyer,
  Cleve, Bennett}. The blackbox in this algorithm has always
previously been described as evaluating a function over the integers
rather than approximating a continuous function with finite
precision. Gradient finding is the first known practical variant
of this algorithm. In \cite{Mosca}, the question as to whether the
algorithm could be adapted for any task of practical interest was
presented as an open problem, which this paper resolves.

The blackbox that we consider takes as its input $d$ binary strings,
each of length $n$, along with $n_o$ ancilla qubits which should be
set to zero. The blackbox writes its output into the ancilla bits
using addition modulo $2^{n_o}$ and preserves the input bits. This is
a standard technique for making any function reversible, which it must
be for a quantum computer to implement it.

The purpose of the blackbox is to evaluate some function $f:
\mathbb{R}^d \to \mathbb{R}$ with $n_o$ bits of precision on a finite
domain. It does so in fixed-point notation. That is, the inputs and
outputs to the function $f$, which are real numbers within a finite
range, are approximated by the inputs and outputs to the blackbox,
which are integers ranging from 0 to $2^{n}$, and 0 to $2^{n_o}$,
respectively, via appropriate scaling and offset. 

For numerical gradient estimation to work, in the classical or quantum
case, $f$ and its first partial derivatives must be continuous in the
vicinity of the point $(x_1,x_2,\ldots,x_d)$ at which the gradient is
to be evaluated. Classically, to estimate $\nabla f$ at $\bx$ in $d$
dimensions, one can evaluate $f$ at $\bx$ and at $d$ additional
points, each displaced from $f$ along one of the $d$ dimensions.

In practice, it may be desirable in the classical gradient estimation
algorithm to perform the function evaluations displaced by $\pm l/2$
from $\bx$ along each dimension so that 
the region being sampled is centered at $\bx$. In this case $2d$
function evaluations are required instead of $d+1$.
$\partial f/\partial x_1$ will then be given by
$(f(x_1+l/2,\ldots)-f(x_1-l/2,\ldots))/l$, and similarly for the other
partial derivatives. Inserting the Taylor expansion for $f$ about
$\bx$ into this expression shows that the quadratic terms will cancel,
leaving an error of order $l^2$ and higher. One must choose $l$
sufficiently small that these terms are negligible.

Now we consider the quantum case. It suffices to show how to perform a
quantum gradient estimation at $\bx=\bo$,
since the gradient at other points can be obtained by trivially
redefining $f$. To estimate the gradient at the origin, start with $d$
input registers of $n$ qubits each, plus a single output register of
$n_o$ qubits, all initialized to zero. Perform the Hadamard transform
on the input registers, write the value 1 into the output register and
then perform an inverse Fourier transform on it. This yields the
superposition

\[
\frac{1}{\sqrt{N^d N_o}} \sum_{\delta_1=0}^{N-1} \sum_{\delta_2=0}^{N-1}
\ldots \sum_{\delta_d=0}^{N-1}\ket{\delta_1} \ldots \ket{\delta_d}
\sum_{a=0}^{N_o-1} e^{i 2 \pi a/N_o} \ket{a}
\]

where $N \equiv 2^n$ and $N_o \equiv 2^{n_o}$. In vector notation,

\[
=\frac{1}{\sqrt{N^d N_o}} \sum_{\bd} \ket{\bd} \sum_a e^{i 2 \pi a/N_o}
\ket{a}
\]

Next, use the blackbox to compute $f$ and add it modulo $N_o$ into the
output register. The output register is an eigenstate of addition
modulo $N_o$. The eigenvalue corresponding to addition of x is $e^{i 2
  \pi x/N_o}$. Thus by writing into the output register via modular
addition, we obtain a phase proportional to $f$. This technique is
sometimes called phase kickback. The resulting state is

\[
\frac{1}{\sqrt{N^d N_o}} \sum_{\bd}
e^{i 2 \pi \frac{N}{m l} f(\frac{l}{N}(\bd-\frac{\bN}{2}))}
\ket{\bd} \sum_a e^{i 2 \pi a/N_o}\ket{a}
\]

where $\bN$ is the $d$-dimensional vector $(N,N,N,\ldots)$, and $l$ is
the size of the region over which $f$ is approximately linear. $l$ and
$\bN$ are used to convert from the components of $\bd$, which are
nonnegative integers represented by bit strings, to rationals evenly
spaced over a small region centered at the origin. Similarly, the
blackbox output is related to the value of $f$ by $a \to a \oplus
\lceil \frac{N N_o}{ml} f \rfloor \mod N_o$. $m$ is the size of the
interval which bounds the components of $\nabla f$. This ensures
proper scaling of the final result into a fixed point representation,
that is, as an integer from 0 to $2^n$.

For sufficiently small $l$,

\[
\approx \frac{1}{\sqrt{N^d N_o}} \sum_{\bd}
e^{i \frac{2 \pi N}{m l} \left( f(\bo) +
  \frac{l}{N}(\bd-\frac{\bN}{2}) \cdot \nabla f \right)} \ket{\bd}
\sum_a e^{i 2 \pi a/N_o} \ket{a}.
\]

Writing out the vector components, and ignoring global phase, the
input registers are now approximately in the state 

\[
\begin{split}
= \frac{1}{\sqrt{N^d}} \sum_{\delta_1 \ldots
  \delta_{d}} 
e^{i \frac{2 \pi}{m} \left( \delta_1 \frac{\partial
    f}{\partial x_1} + \delta_2 \frac{\partial f}{\partial x_2} +
  \ldots + \delta_{d} \frac{\partial f}{\partial x_{d}} \right)}
\times \\
\ket{\delta_1} \ket{\delta_2} \ldots \ket{\delta_{d}}.
\end{split}
\]

This is a product state:

\[
=\frac{1}{\sqrt{N^d}} \left( \sum_{\delta_1} e^{i
  \frac{2 \pi}{m} \delta_1 \frac{\partial f}{\partial x_1}}
  \ket{\delta_1} \right) \ldots \left( \sum_{\delta_d} e^{i \frac{2
  \pi}{m} \delta_d \frac{\partial f}{\partial x_d}} \ket{\delta_d}
  \right).
\]

Fourier transform each of the registers, obtaining

\[
 \Ket{\frac{N}{m} \frac{\partial f}{\partial
  x_1}} \Ket{\frac{N}{m} \frac{\partial f}{\partial x_2}} \ldots
  \Ket{\frac{N}{m} \frac{\partial f}{\partial x_d}}.
\]

Then simply measure in the computational basis to obtain the
components of $\nabla f$ with $n$ bits of precision. Because $f$ will
in general not be perfectly linear, even over a small region, there
also will be nonzero amplitude to measure other values close to the
exact gradient, as will be discussed later.

Normally, the quantum Fourier transform is thought of as mapping the
discrete planewave states to the computational basis states:

\[
\frac{1}{\sqrt{N}}\sum_{j=0}^N e^{2 \pi ijk/N} \ket{j} \to \ket{k}
\]

where $0 < k < N$. However, negative $k$ is also easily dealt with,
since

\[
\frac{1}{\sqrt{N}}\sum_{j=0}^N e^{-2 \pi ij|k|/N} \to \ket{N-|k|}.
\]
 
Thus negative components of $\nabla f$ pose no difficulties for the
quantum gradient estimation algorithm provided that bounds for the
values of the components are known, which is a requirement for any
algorithm using fixed-point arithmetic.

In general the number of bits of precision necessary to represent a
set of values is equal to $\log_2 (r/\delta)$, where $r$ is the range
of values, and $\delta$ is the smallest difference in values one
wishes to distinguish. Thus for classical gradient estimation with $n$
bits of precision, one needs to evaluate $f$ to

\begin{equation}
\label{classprec}
\log_2 \left[ \frac{\mathrm{max}(f) -
    \mathrm{min}(f)}{\frac{ml}{2^n}}\right] 
\end{equation}

bits of precision.

An important property of the quantum Fourier transform is that it can
correctly distinguish between exponentially many discrete planewave
states with high probability without requiring the phases to be
exponentially precise \cite{Nielsen}. It is not hard to show that if
each phase is accurate to within $\theta$ then the inner product
between the ideal state and the actual state is at least $\cos \theta$,
and therefore the algorithm will still succeed with probability at
least $\cos^2 \theta$.

As shown earlier, the phase acquired by ``kickback'' is equal to
$\frac{2 \pi N}{ml} f$, and therefore, for the phase to be accurate
to within $\pm \theta$, $f$ must be evaluated to within $\pm
\frac{ml}{2 \pi N} \theta$. Thus, recalling that $N=2^n$,

\begin{equation}
\label{quantprec}
n_o = \log_2 \left[ \frac{\max f - \min f}{\frac{ml}{2^n}
    \frac{\theta}{2 \pi}} \right].
\end{equation}

As an example, if $\theta = \pi/8$, then the algorithm will behave
exactly as in the idealized case with approximately $85\%$ probability,
and $N_o$ will exceed the classically required precision by
four bits, for a given value $l$. $l$ also differs between the quantum and
classical cases, as will be discussed later. Thus $n_o$ differs from the
classically required precision only by an additive constant which
depends on $\theta$ and $l$. Because the classical and quantum precision
requirements are both proportional to $n$, this difference becomes
negligible in the limit of large $n$.

The only approximation made in the description of the quantum gradient
estimation algorithm was expanding $f$ to first order. Therefore the
lowest order error term will be due to the quadratic part of $f$. The
behavior of the algorithm in the presence of such a quadratic term
provides an idea of its robustness. Furthermore, in order to minimize
the number of bits of precision to which $f$ must  be evaluated, $l$
should be chosen as large as possible subject to the constraint that
$f$ be locally linear. The analysis of the quadratic term provides a
more precise description of this constraint.

The series of quantum Fourier transforms on different registers can be
thought of as a single $d$-dimensional quantum Fourier
transform. Including the quadratic term, the state which this Fourier
transform is acting on has amplitudes

\[
f(\bd)=\frac{1}{N^{d/2}} \mathrm{exp}\left[ i 2 \pi \left( \bd
  \cdot \nabla f+\frac{l}{2 m N} \bd^T H \bd \right) \right],
\]

where $H$ is the Hessian matrix of $f$. After the Fourier transform,
the amplitudes should peak around the correct value of $\nabla
f$. Here we are interested in the width of the peak, which should not
be affected by $\nabla f$, so for simplicity it will be set to 0. The
Fourier transform will yield amplitudes of \footnote{$\bd$ here really
  represents $\bd-\bN/2$.}

\[
\tilde{f}(\bk)=\frac{1}{N^d} \sum_{\bd} \mathrm{exp} \left[i 2 \pi
  \left( \frac{l}{2 m N} \bd^T H \bd-\frac{1}{N} \bk \cdot \bd \right)
  \right].
\]

Ignoring global phase and doing a change of variables
($\bu=\delta/N$),

\[
\approx \int_{-1/2}^{1/2} \ldots \int_{-1/2}^{1/2} \mathrm{exp} \left[
  i 2 \pi \left( \frac{Nl}{2m} \bu^T H \bu - \bk \cdot \bu \right)
  \right] \ud^d \bu
\]

This integral can be approximated using the method of stationary
phase.
$
\nabla \phi = \frac{Nl}{2m} \left( H^T+H \right) \bu - \bk
$
but Hessians are symmetric, so
$
\nabla \phi = \frac{N l}{m} H \bu-\bk.
$
Thus (again ignoring global phase),

\[
\tilde{f}(\bk)  \approx \left\{ \begin{array}{ll} 
\sqrt{\frac{1}{\mathrm{Det}\left( \frac{Nl}{m} H \right)}} &
\textrm{if}\ \exists\ \bu\in C\ \textrm{s.t.}\ \frac{Nl}{m}H\bu-\bk=0\\ 
\ & \ \\
0 & \textrm{otherwise}
\end{array}
\right.
\]

where $C$ is the region $-1/2 < u_i < 1/2 \quad \forall\ i$. So
according to the stationary phase approximation, the peak is simply a
region of uniform amplitude, with zero amplitude
elsewhere. Geometrically, this region is what is obtained by applying 
the linear transformation $\frac{Nl}{m}H$ to the $d$-dimensional unit
hypercube. 

Since we have set $\nabla f = 0$, the variance of $\frac{N}{m}
\frac{\partial f}{\partial x_i}$ will be

\[
\sigma_i^2=\frac{1}{\mathrm{Det} A} \int_D k_i^2 d^d \bk \quad
\textrm{where} \quad A= \frac{Nl}{m} H
\]

and $D$ is the region of nonzero amplitude. Doing a change of
variables with $A$ as the Jacobian, 

\[
\sigma_i^2 = \frac{1}{\mathrm{Det} A} \int_C (A \bk')_i^2
\mathrm{Det} A\ d^d\bk'
\]

where $C$ is again the unit hypercube centered at the origin. In
components,

\[
\sigma_i^2 = \int_C \left( \sum_j A_{ij} k_j' \right)^2 d^d \bk'.
\]

The expectation values on a hypercube of uniform probability are
$\left< x_i x_j \right> = \frac{1}{12} \delta_{ij}$, thus

\[
\sigma_i^2=\frac{1}{12} \sum_j A_{ij}^2=\frac{N^2 l^2}{12 m^2} \sum_j
  \left( \frac{\partial^2 f}{\partial x_i \partial x_j} \right)^2.
\] 

This quadratic dependence on $N$ is just as expected since, at the end
of the computation, the register that we are measuring is intended to
contain $\frac{N}{m} \frac{\partial f}{\partial x}$. Therefore the
uncertainty in $\partial f/\partial x_i$ is approximately

\begin{equation}
\label{statphase}
\frac{l}{2 \sqrt{3}} \sqrt{\sum_j \left( \frac{\partial^2 f}{\partial
    x_i \partial x_j} \right)^2}
\end{equation}

independent of $N$. In the classical algorithm which uses $2d$
function evaluations, the cubic term introduces an error of $\sigma
\sim \frac{l^2}{24} D_3$ where $D_3$ is the typical
\footnote{Alternatively, we can define $D_3$ and $D_2$ as the largest
  $2^{\textrm{nd}}$ and $3^{\textrm{rd}}$ partial derivatives of $f$
  to obtain a worst case requirement on $l$.} magnitude of third
partial derivatives of $f$. If the  $2^{\textrm{nd}}$ partial
derivatives of $f$ have a magnitude of approximately $D_2$ then the
typical uncertainty in the quantum case will be $\sigma \sim \frac{l
D_2 \sqrt{d}}{2 \sqrt{3}}$. To obtain a given uncertainty $\sigma$, 

\[
l  \sim \left\{ \begin{array}{ll} 
2 \sqrt{\frac{6 \sigma}{D_3}}  & \textrm{classical}\\ 
\ & \ \\
\frac{2 \sqrt{3} \sigma}{D_2 \sqrt{d}} & \textrm{quantum}
\end{array}
\right.
\]

Recalling Eq. (\ref{classprec}) and (\ref{quantprec}), the number of
bits of precision to which $f$ must be evaluated depends
logarithmically on $l$. However, in the limit of large $n$, the number
of bits will match the classical requirement.

\begin{figure}
\includegraphics[width=0.36\textwidth]{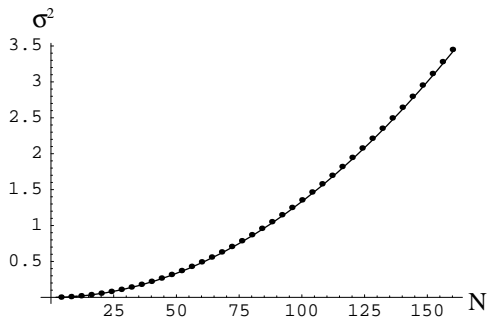}
\caption{\label{ndeplab} Comparison between error estimates obtained in the
  stationary phase approximation (solid line) and numerical results
  (points) for the one dimensional case. Here the $2^{\textrm{nd}}$
  derivative remains constant ($\alpha=0.02$), and the number of bits
  to which the gradient is being evaluated is varied.}
\end{figure}

\begin{figure}
\includegraphics[width=0.36\textwidth]{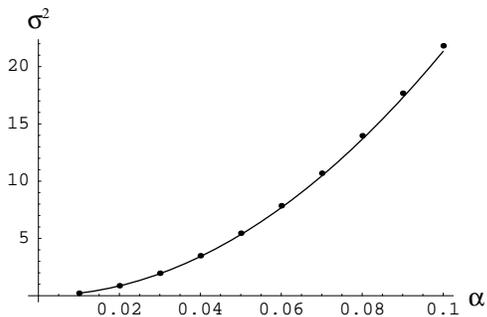}
\caption{\label{ddeplab} Analytical error estimates (solid line) are
  again tested against numerical results (the points) in the one
  dimensional case, varying the $2^{\textrm{nd}}$ derivative instead
  of the number of bits. Here $N=80$.}
\end{figure}

The level of accuracy of the stationary phase approximation can be
assessed by comparison to numerical solutions of example cases. In one
dimension, Eq. (\ref{statphase}) reduces to
$\sigma^2 = \frac{\alpha^2 N^2}{3}$ where $\alpha=\frac{l}{2 m}
\frac{\partial^2 f}{\partial x^2}$. Figures \ref{ddeplab} and
\ref{ndeplab} display the close agreement between numerical results
and the analytical solution obtained using stationary phase.

\begin{figure}
\includegraphics[width=0.45\textwidth]{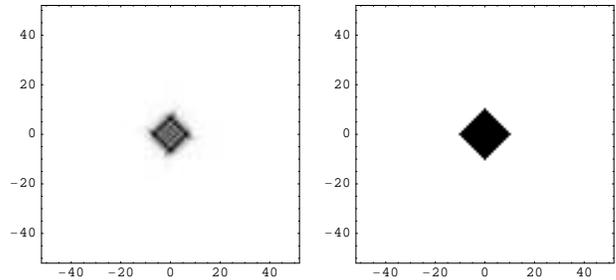}
\caption{\label{peakshapes} On the left, the probability, as numerically
  calculated, is shown. The areas of highest probability appear
  darkest. On the right, the region of nonzero probability, as
  calculated in the stationary phase approximation, is shaded in
  black.}
\end{figure}

A two dimensional example provides a nontrivial test of the
stationary phase method's prediction of the peak shape. If the Hessian
is such that

\[
\frac{N}{m} H = 0.1 \left[ \begin{array}{rr}
  1 & 1 \\
  1 & -1
\end{array}
\right]
\]

then, according to the stationary phase approximation, the peak should
be a square of side length $\frac{\sqrt{2}}{10} l$ with a $45^\circ$ 
rotation. This is in reasonable agreement with the numerical result,
as shown in figure \ref{peakshapes}.

Because this algorithm requires only one blackbox query, one might
expect that it could be run recursively to efficiently obtain higher
derivatives. In this case, another instance of the same algorithm
serves as the blackbox. However, the algorithm itself differs from the
blackbox in that the blackbox has scalar output which it adds modulo
$N_o$ to the existing value in the output register, and it does not
incur any input-dependent global phase. An additive scalar output can
be obtained by minor modification to this algorithm, but the most
straightforward techniques for eliminating the global phase require an
additional blackbox query, thus necessitating $2^n$ queries for the
evaluation of an $n^{th}$ partial derivative, just as in the classical
case. 

The problem of global phase when recursing quantum algorithms as well
as the difficulties inherent in recursing approximate or probabilistic
algorithms are not specific to gradient finding but are instead fairly
general.

Efficient gradient estimation may be useful, for example, in some
optimization and rootfinding algorithms. Furthermore, upon
discretization, the problem of minimizing a functional is converted
into the problem of minimizing a function of many variables, which
might benefit from gradient descent techniques. A speedup in the
minimization of functionals may in turn enable more efficient solution
of partial differential equations via the Euler-Lagrange equation.
The analysis of the advantage which this technique can provide in
quantum numerical algorithms remains open for further research. 

The author thanks P. Shor, E. Farhi, L. Grover, J. Traub, and
M. Rudner for useful discussions, and MIT's Presidential Graduate
Fellowship program for financial support.

\bibliography{gradient}

\end{document}